%% file: main.tex
\input{templates/layout}
\input{templates/article_setup}

\title{WebCraftBench: Evaluating Web Application Generation from a Software Testing Perspective}

\author[1,\textdagger]{Chenxu~Liu}
\author[1]{Zilu~Zou}
\author[2,1]{Peizhong~Gao}
\author[3,1]{Jiawen~Tao}
\author[2,1]{Zhexin~Zhang}
\author[1]{Guang~Chen}
\author[1]{Haowei~Lin}
\author[1]{Ying~Zhou}
\author[1]{Tianyi~Bai}
\author[1]{Dolly~Deng}
\author[1]{Suncong~Zheng}
\author[1]{Maxm~Pan}

\affil[1]{Hunyuan Team, Tencent}
\affil[2]{Tsinghua University}
\affil[3]{Peking University}

\authoremails{%
  \{lewiscxliu, ziluzou, chillgao, jadentao, zhexinzhang, maxluxchen, cloudlin,\\
  lilithyzhou, tianabai, dollydeng, congzheng, maxmpan\}@tencent.com}

\correspondingauthor{Chenxu Liu\textsuperscript{\textdagger} (lewiscxliu@tencent.com)}

\begin{document}

\input{sections/paper}

\end{document}

%% file: templates/layout.tex
\PassOptionsToPackage{framemethod=tikz}{mdframed}
\documentclass[11pt,letterpaper,onecolumn,logo]{templates/paper}
\graphicspath{{figures/}{templates/assets/}}

\usepackage[authoryear,round]{natbib}
\usepackage[most,breakable,skins]{tcolorbox}
\tcbuselibrary{skins,breakable}
\usepackage{microtype}

\let\cite\citep

\input{templates/common}

\AtBeginDocument{%
  \setlength{\parindent}{0pt}%
  \setlength{\parskip}{0.5\baselineskip plus 2pt}%
}

\makeatletter
\renewcommand{\abscontent}{}
\renewenvironment{abstract}
  {\begin{tcolorbox}[
      colback=gblue9!5,
      colframe=gblue9!5,
      boxrule=0pt,
      arc=6pt,
      left=8pt,right=8pt,
      top=6pt,bottom=6pt,
      breakable]
   \absfont}
  {\end{tcolorbox}\par\bigskip}
\makeatother

%% file: templates/common.tex
\usepackage{amsmath,amsfonts,amssymb,amsthm}
\usepackage{booktabs,array,tabularx,multirow,ragged2e,xcolor}
\usepackage{tikz}
\usetikzlibrary{arrows.meta,positioning,shapes.geometric,fit,backgrounds,calc}
\usepackage[most]{tcolorbox}

\definecolor{codeBg}{HTML}{F8FAFC}
\definecolor{codeFrame}{HTML}{D5DDE8}
\definecolor{codeTitleBg}{HTML}{EEF4FF}
\definecolor{codeTitleText}{HTML}{1E3A8A}
\definecolor{codeAccent}{HTML}{2563EB}
\definecolor{tableHeadBg}{HTML}{EAF1F8}
\definecolor{tableBandBg}{HTML}{F7FAFC}
\definecolor{tableGroupBg}{HTML}{F1F5F9}
\definecolor{tableRule}{HTML}{94A3B8}
\definecolor{tableText}{HTML}{1F2937}

\usepackage{enumitem}
\setlist[itemize]{leftmargin=*}
\setlist[enumerate]{leftmargin=*}

\theoremstyle{definition}

\AtBeginEnvironment{tabular}{\arrayrulecolor{tableRule}}
\AtBeginEnvironment{tabularx}{\arrayrulecolor{tableRule}}

\newtcblisting{boxedcode}[1][]{
  enhanced,
  listing only,
  listing engine=listings,
  breakable,
  width=0.88\textwidth,
  center,
  colback=codeBg,
  colframe=codeFrame,
  colbacktitle=codeTitleBg,
  coltitle=codeTitleText,
  fonttitle=\sffamily\bfseries\footnotesize,
  boxrule=0.45pt,
  arc=2pt,
  borderline west={2pt}{0pt}{codeAccent},
  left=9pt,
  right=9pt,
  top=7pt,
  bottom=7pt,
  before skip=8pt,
  after skip=10pt,
  listing options={
    basicstyle=\ttfamily\scriptsize\color{black!82},
    breaklines=true,
    columns=fullflexible,
    keepspaces=true,
    showstringspaces=false
  },
  #1
}

%% file: templates/article_setup.tex
\input{templates/math_commands}

\usepackage{float}
\usepackage{titletoc}
\setcitestyle{authoryear,round,citesep={;},aysep={,},yysep={;}}
\hypersetup{hypertexnames=false}

\definecolor{zpositive}{RGB}{121,174,144}
\definecolor{znegative}{RGB}{194,132,143}
\newcommand{\zposcell}[2]{\cellcolor{zpositive!#1}\ensuremath{#2}}
\newcommand{\znegcell}[2]{\cellcolor{znegative!#1}\ensuremath{#2}}

\newtcolorbox{promptbox}[1][]{
  enhanced,
  breakable,
  colback=black!2,
  colframe=black!35,
  colbacktitle=black!6,
  coltitle=black,
  boxrule=0.5pt,
  arc=1.5pt,
  title=#1,
  fonttitle=\bfseries\small,
  before upper={\ttfamily\scriptsize\raggedright\setlength{\parindent}{0pt}},
  left=5pt,
  right=5pt,
  top=4pt,
  bottom=4pt,
  before skip=6pt,
  after skip=8pt
}

\newcommand{\numtasks}{369}

%% file: templates/math_commands.tex
\usepackage{amsmath,amsfonts,bm}

\def\eqref#1{equation~\ref{#1}}

\def\1{\bm{1}}

\DeclareMathAlphabet{\mathsfit}{\encodingdefault}{\sfdefault}{m}{sl}
\SetMathAlphabet{\mathsfit}{bold}{\encodingdefault}{\sfdefault}{bx}{n}



%% file: sections/paper.tex
\maketitle

\begin{abstract}
Human evaluation provides a direct measure of the quality of LLM-generated web applications. However, fitting human judgments through automated evaluation remains challenging. Static benchmarks can credit functionality that exists in source code but is unreachable at runtime. Interactive benchmarks exercise the application, yet incomplete exploration can cause them to miss implemented functionality and confound application defects with agent execution failures.
To address these limitations, we propose WebCraftBench, an interactive benchmark for evaluating web application generation from a software testing perspective. WebCraftBench instruments each generated application and uses code coverage to guide an agent in exploring its functionality through user-simulated interactions. It then abstracts the interaction trace into a state-transition graph and evaluates the application along three dimensions: visual aesthetics, usability, and requirement alignment. By separating exploration from scoring, WebCraftBench collects runtime evidence without constraining exploration to predefined acceptance criteria.
WebCraftBench comprises 369 real-world user requirements and 5,088 acceptance criteria. Evaluation of 17 frontier LLMs reveals distinct strengths across the three dimensions, with no model leading on every dimension. On 197 validated sessions sampled from an internal arena, WebCraftBench achieves 85.3\% agreement with human preferences, with agreement generally increasing as the score difference between paired applications grows. Further experiments show that coverage guidance improves exploration coverage and the model rankings remain stable when the judge model is replaced.
\end{abstract}

\input{sections/rank_teaser}

\section{Introduction}
\label{sec:intro}

Web applications run across platforms without installation, making them an accessible medium for LLM agents to deliver interactive functionality to users. Given a natural-language requirement, an agent iteratively invokes tools and generates the source code of a runnable application. However, the evaluation of generated applications is typically highly subjective. Human-preference leaderboards such as Code Arena~\citep{WebDevArena} obtain user preferences directly by asking annotators to choose between two applications generated by different models for the same requirement. However, repeated human evaluation is costly, requiring an automated benchmark to reduce the cost of evaluation while preserving user preferences.

Existing automated benchmarks face two major limitations, making them less likely to align with user preferences. \textbf{First, static benchmarks do not exercise the generated application.} Static benchmarks based on source code or a single rendered screenshot~\citep{fullfront,webcoderbench} assess the applications without verifying whether users can reach and use the intended functionalities. For example, a broken button on the entry page can block all downstream functionalities, while the corresponding code may still be present, causing the benchmark to credit functionalities that are unusable in practice. \textbf{Second, interactive benchmarks struggle to distinguish application defects from agent execution failures.} Interactive benchmarks define rubrics, reference answers, or contracts and ask an agent to validate them through interactions~\citep{webgenbench,frontalk,miniappbench,webgamebench,webrise,webcompass}. However, a failed check may indicate that the application lacks the target functionality or that the agent failed to locate or trigger it. Errors in GUI interaction and evidence interpretation make these causes difficult to distinguish, introducing uncertainty into the final score.

To address these limitations, in this paper, we propose WebCraftBench, an interactive benchmark for evaluating web application generation from a software testing perspective. WebCraftBench treats each generated application as software under test: it simulates user interactions and explores the application to collect runtime evidence. Code coverage serves two purposes in this process. It guides the exploration toward unexecuted code and quantifies how much of the implementation has been exercised. WebCraftBench separates exploration from scoring so that the exploration process is not constrained by acceptance criteria.

WebCraftBench consists of four stages.
\emph{(i) Instrumentation.} An automated tool instruments the JavaScript code in each generated application to collect runtime code coverage without manual adaptation. 
\emph{(ii) Coverage-guided exploration.} An LLM-driven agent explores the application as a software tester. When coverage plateaus, a separate LLM analyzes the uncovered code and provides natural-language guidance for subsequent exploration. This feedback supplements the agent's own assessment of what remains to be tested. 
\emph{(iii) State abstraction.} A deterministic state abstraction approach processes each page and merges pages with equivalent normalized representations into the same state, generating a compact state-transition graph to represent the exploration trace.
\emph{(iv) Scoring.} LLM and agentic judges assess visual aesthetics, usability, and requirement alignment using the state-transition graph and the recorded traces, DOM snapshots, and screenshots.

We evaluate 17 frontier LLMs on WebCraftBench and validate the benchmark against human preferences from pairwise comparison samples. WebCraftBench achieves 85.3\% pairwise agreement on 197 validated samples. The model-level results reveal complementary strengths in visual aesthetics, usability, and requirement alignment. Additional experiments show that coverage guidance improves exploration coverage and that model rankings remain stable across judge models. As shown in Figure~\ref{fig:frontend-alignment}, WebCraftBench rankings are highly correlated with Code Arena rankings\footnote{Code Arena Frontend, September 10, 2026: \url{https://arena.ai/leaderboard/code/webdev/frontend}.}, with a Spearman rank correlation~\citep{spearman} of $\rho=0.890$ across the 17 shared models. The two evaluations show broadly consistent capability tiers at shared-set ranks 1--5, 6--10, 11--14, and 15--17. Despite differences in reasoning effort and harness configurations, this alignment suggests that WebCraftBench captures capability distinctions consistent with the community preferences reflected in Code Arena.

In summary, this paper makes the following main contributions:
\begin{itemize}
  \item We construct WebCraftBench, a benchmark with \numtasks{} real-world user requirements and 5,088 acceptance criteria for evaluating web application generation.
  \item We design a four-stage evaluation pipeline that separates exploration from scoring, guides exploration with code coverage, and organizes runtime evidence through state abstraction.
  \item We evaluate 17 frontier LLMs across visual aesthetics, usability, and requirement alignment. Validation shows 85.3\% agreement with human preferences, stable rankings across judge models, reliable instrumentation, and improved exploration coverage.
\end{itemize}

\section{Related Work}
\label{sec:related}

\subsection{Benchmarks for Web Application Generation}
Existing related benchmarks are shown in Table~\ref{tab:related-work}. Early static benchmarks formulate web generation as reproducing a reference design and measure pixel-, text-, or element-level similarity to a reference screenshot or implementation~\citep{pix2code,websight,design2code,fullfront}. Real-world requirements, however, often admit multiple valid implementations. WebCoderBench~\citep{webcoderbench} assesses source code and initial renderings without a unique reference implementation. These benchmarks support inexpensive, reproducible evaluation, but do not exercise the application and may credit functionality that is unreachable at runtime.

Interactive benchmarks execute applications in a browser, and score them using test suites, fixed action sequences, or staged screenshots~\citep{webbench,frontendbench,artifactsbench}, or ask agents to validate checklists, reference answers, contracts, or specifications~\citep{webgenbench,miniappbench,webrise,webgamebench,frontalk}. These benchmarks observe a broader range of runtime behavior and provide evidence for failure diagnosis. Predefined targets, nevertheless, constrain the verification scope, leaving behavior outside those targets unassessed. Failed agent-based checks can also reflect either application defects or agent failures.

More recently, Cookie-Bench~\citep{cookiebench} removes checklists, allows the agent to plan interactions from the user request, and scores applications after evidence collection. I-WebGenBench~\citep{iwebgenbench} applies standardized actions to enumerated components and records DOM changes. However, neither explicitly measures exploration adequacy. WebCraftBench introduces code coverage measurement into the exploration process to quantify the functionalities exercised and guide further exploration, while withholding acceptance criteria from the exploration agent.

\begin{table*}[t]
\caption{Comparison of WebCraftBench with existing benchmarks for web application generation.}
\label{tab:related-work}
\setlength{\tabcolsep}{1pt}
\resizebox{\textwidth}{!}{
\begin{tabular}{lcccccccc}
\toprule
\multirow{2}{*}{\textbf{Benchmark}} &
\multirow{2}{*}{\textbf{Samples}} &
\multirow{2}{*}{\textbf{Interactive}} &
\multirow{2}{*}{\textbf{Exploration}} &
\multicolumn{3}{c}{\textbf{Judgment}} &
\multirow{2}{*}{\shortstack{\textbf{Exploration--Scoring}\\\textbf{Separation}}} &
\multirow{2}{*}{\shortstack{\textbf{Exploration}\\\textbf{Quantification}}} \\
\cmidrule(lr){5-7}
 & & & & \textbf{Rules} & \textbf{LLM} & \textbf{Manual} & & \\ \midrule
FullFront~\citep{fullfront}           & 400  & $\times$   & None                  & \checkmark & \checkmark & $\times$   & $\times$    & $\times$ \\
WebCoderBench~\citep{webcoderbench}   & 1,572  & $\times$   & None                  & \checkmark & \checkmark & $\times$   & $\times$    & $\times$ \\
FrontendBench~\citep{frontendbench}   & 148    & \checkmark & Scripted              & \checkmark & $\times$   & $\times$   & $\times$    & $\times$ \\
Web-Bench~\citep{webbench}            & 1,000    & \checkmark & Scripted              & \checkmark & $\times$   & $\times$   & $\times$    & $\times$ \\
ArtifactsBench~\citep{artifactsbench} & 1,825  & \checkmark & Scripted              & $\times$   & \checkmark & $\times$   & \checkmark  & $\times$ \\ \midrule
WebDev Arena~\citep{WebDevArena}      & --     & \checkmark & Human                 & $\times$   & $\times$   & \checkmark & $\times$    & $\times$ \\
Design Arena~\citep{DesignArena}      & --     & \checkmark & Human                 & $\times$   & $\times$   & \checkmark & $\times$    & $\times$ \\ \midrule
WebGen-Bench~\citep{webgenbench}      & 101    & \checkmark & Checklist-guided      & $\times$   & \checkmark & $\times$   & $\times$    & $\times$ \\
FronTalk~\citep{frontalk}             & 100    & \checkmark & Checklist/autonomous  & $\times$   & \checkmark & $\times$   & $\times$    & $\times$ \\
MiniAppBench~\citep{miniappbench}     & 500    & \checkmark & Checklist-guided      & $\times$   & \checkmark & $\times$   & $\times$    & $\times$ \\
WebCompass~\citep{webcompass}         & 1,526  & \checkmark & Checklist-guided      & $\times$   & \checkmark & $\times$   & $\times$    & $\times$ \\
WebGameBench~\citep{webgamebench}     & 111    & \checkmark & Specification-guided  & $\times$   & \checkmark & $\times$   & $\times$    & $\times$ \\
WebRISE~\citep{webrise}               & 442    & \checkmark & Contract-guided       & \checkmark & \checkmark & $\times$   & $\times$    & $\times$ \\ \midrule
I-WebGenBench~\citep{iwebgenbench}    & 201    & \checkmark & Enumerative           & \checkmark & \checkmark & $\times$   & $\triangle$ & $\times$ \\
Cookie-Bench~\citep{cookiebench}      & 1,000  & \checkmark & Model-autonomous      & $\times$   & \checkmark & $\times$   & \checkmark  & $\times$ \\ \midrule
WebCraftBench (ours)                     & 369    & \checkmark & Coverage-guided       & $\times$   & \checkmark & $\times$   & \checkmark  & \checkmark \\ \bottomrule
\end{tabular}}

{\footnotesize
\textbf{Note:} $\triangle$ indicates partial satisfaction.
}
\end{table*}

\subsection{Automated GUI Testing}
Automated GUI testing exercises applications through GUI actions to gain coverage and reveal failures within a specific time budget~\citep{empirical}. Existing work studies state abstraction approaches at different granularities~\citep{fraggen,judge,booster} to avoid repeated exploration, while using random-based~\citep{monkey}, model-based~\citep{crawljax}, reinforcement-learning-based~\citep{webexplor}, and LLM-based~\citep{gptdroid,temac} exploration strategies. 

WebCraftBench adapts these approaches to evaluating LLM-generated web applications. Code coverage provides exploration feedback and quantifies the implementation exercised, leveraging the ideology of LLM-based GUI testing approaches. State abstraction groups near-duplicate pages~\citep{ndstudy} into a compact state-transition graph, as part of the evidence for scoring.

\section{Dataset}
\label{sec:dataset}

WebCraftBench contains 369 real-world user requirements, each paired with a checklist used exclusively for scoring.

\subsection{Data Source}
\label{sec:data-source}

We collect our user requirements from production traffic on an enterprise-internal replica of Code Arena~\citep{WebDevArena}. These requirements retain characteristics of real-world usage that are difficult to reproduce synthetically, including colloquial expressions, incomplete specifications, and substantial variation in the expected scope of the application. Following WebCoderBench~\citep{webcoderbench}, we first anonymize the requirements. Three annotators then review each requirement and use majority voting to exclude samples that are incomprehensible, depend on unavailable materials, or fall outside the web application scope. We remove lexical duplicates using MinHash~\citep{minhash} and semantic duplicates using MiniLM~\citep{minilm}. The resulting dataset contains 369 text-only requirements. Our requirements have a mean length of 601 characters, a median of 105, and a range of 7--37,903. This distribution captures the coexistence of brief requests and detailed specifications in production traffic.

\subsection{Requirement Characterization}
\label{sec:data-taxonomy}

Following the annotation protocol of WebCoderBench~\citep{webcoderbench}, we characterize each requirement along four dimensions: artifact complexity, expression style, requirement clarity, and application category. An LLM assigns initial labels and provides a rationale for each, after which human annotators review and revise the annotations. Table~\ref{tab:dataset-stats} summarizes the results. Overall, 63.4\% of requirements are not fully specified ($C2$ or $C3$), and 60.2\% use colloquial language ($S2$). These characteristics require LLMs to interpret user intent from incomplete or informal descriptions. Entertainment and scientific demonstrations are the two largest application categories.

\begin{table}[t]
\caption{Dataset statistics for WebCraftBench.}
\label{tab:dataset-stats}
\centering
\resizebox{0.8\textwidth}{!}{
\begin{tabular}{llr@{\hskip 1.5em}llr@{\hskip 1.5em}lr}
\toprule
\multicolumn{3}{c}{\textbf{Artifact Complexity}} & \multicolumn{3}{c}{\textbf{Expression Style}} & \multicolumn{2}{c}{\textbf{Application Category}} \\
\cmidrule(r){1-3} \cmidrule(r){4-6} \cmidrule{7-8}
$L1$ & Highly Simple   & 48  & $S1$ & Technical    & 81  & Entertainment       & 106 \\
$L2$ & Simple          & 161 & $S2$ & Colloquial   & 222 & Scientific Demo     & 37  \\
$L3$ & Medium          & 120 & $S3$ & Role-playing & 15  & Online Office Suite & 33  \\
$L4$ & Complex         & 27  & $S4$ & Analogy      & 51  & Utility Website     & 29  \\
$L5$ & Highly Complex  & 13  & \multicolumn{3}{c}{} & Data Visualization  & 21  \\
\cmidrule(r){1-3} \cmidrule(r){4-6}
\multicolumn{3}{c}{\textbf{Requirement Clarity}} & \multicolumn{3}{c}{\textbf{Acceptance Criteria}} & Online Education    & 18  \\
\cmidrule(r){1-3} \cmidrule(r){4-6}
$C1$ & Clear         & 135 & \multicolumn{2}{l}{Functional} & 2,706 & AI-powered      & 15  \\
$C2$ & Intermediate  & 134 & \multicolumn{2}{l}{Content}    & 1,447 & Multimedia      & 15  \\
$C3$ & Vague         & 100 & \multicolumn{2}{l}{Visual}     & 935   & \emph{12 others} & 95  \\
\bottomrule
\end{tabular}}
\end{table}

\subsection{Acceptance Criteria}
\label{sec:data-gt}

Real-world user requirements often admit multiple valid implementations. We therefore associate each task with a checklist, including three sets of \emph{acceptance criteria}, with each acceptance criterion expressing a concise, high-level property that should be observable in the delivered application. The three sets of criteria correspond to three dimensions: \textbf{functional}, including features and interactions; \textbf{content}, including text, data, and specified resources; and \textbf{visual}, including layout, style, and required components. These dimensions jointly determine the requirement-alignment score (detailed in \S\ref{sec:eval-score}). Following WebCoderBench~\citep{webcoderbench}, we use Claude-Opus-4.8, GPT-5.6-Sol, and Grok-4.5 to generate checklists independently. Claude-Opus-4.8 merges the checklists following the majority voting process, and human experts validate the resulting set. The checklists describe the required observable properties without prescribing implementation details or introducing requirements absent from the original request. The 369 requirements yield 5,088 acceptance criteria: 2,706 functional, 1,447 content, and 935 visual. Each task has 13.8 criteria on average, ranging from 2 to 77 according to the detail of the requirement.

\section{Evaluation Process}
\label{sec:eval}

\begin{figure}[t]
\centering
\includegraphics[width=\textwidth]{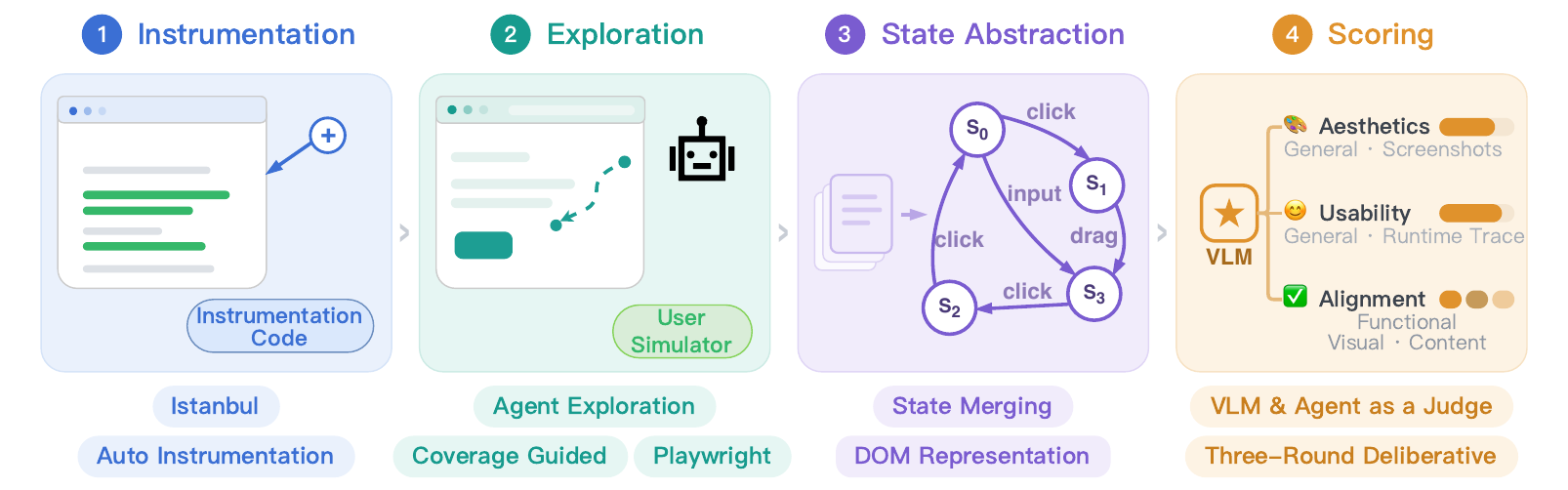}
\caption{The evaluation workflow of WebCraftBench.}
\label{fig:pipeline}
\end{figure}

The evaluation workflow of WebCraftBench contains four stages, which are instrumentation, exploration, state abstraction, and scoring (shown in Figure~\ref{fig:pipeline}). The first three stages collect and organize runtime evidence without access to acceptance criteria, while the scoring stage leverages these criteria for a fair judgment.

\subsection{Instrumentation}
\label{sec:eval-instrument}

In order to obtain coverage from various kinds of generated applications, we develop an automated instrumentation tool based on Istanbul~\citep{istanbul}, a mature and famous instrumentation tool for JavaScript. Our tool supports plain static HTML, Vite, Next, Create React App (CRA), and Astro, with framework-specific procedures applied automatically.
With an application as input, our tool first detects its framework according to configuration files. For static frameworks, it runs Istanbul locally to instrument the application; for frameworks that require a build, it injects a plugin during the build and performs instrumentation after the build. After instrumentation, each JavaScript statement is associated with a counter that records every execution of the statement at runtime. During exploration, after each interaction, cumulative coverage and per-step gains can be calculated from the number of covered statement counters and the total number of counters. 
If instrumentation fails, the application remains in the evaluation and undergoes exploration and scoring in black-box mode. Appendix~\ref{app:instrumentation-details} describes framework support and deployment procedures; \S\ref{sec:results-instrumentation} evaluates instrumentation reliability. The instrumented application is then deployed in a sandbox on a remote server and is accessible through a URL.

\subsection{Coverage-Guided Exploration}
\label{sec:eval-explore}

In order to flexibly control the agent's context during exploration, we develop an agent based on the AWorld~\citep{aworld} framework, interacting with the web applications using Playwright MCP~\citep{playwright}. The agent reads a simplified DOM tree of the current page at each step to understand its content and retrieves only the tree diff between steps to save context space. The agent is prompted as a software tester to discover available functionalities and test boundary conditions. The agent has no access to source code, acceptance criteria, or reference answers.

During exploration, the harness monitors the coverage and detects a coverage plateau after three consecutive interactions without a coverage increase. It then extracts the uncovered code and invokes a separate LLM to produce natural-language suggestions. These suggestions are injected into the exploration agent's context to guide further testing. Exploration ends when the agent declares testing complete or exhausts a 100-step budget. To conclude, in its final step, the agent summarizes the functionality, interactions, and observed failures. Coverage thus both quantifies code execution and directs subsequent exploration; \S\ref{sec:results-exploration} evaluates its contribution.

\subsection{State Abstraction}
\label{sec:eval-graph}

Exploration traces, screenshots, and DOM snapshots can easily exceed the judge's context window and obscure relevant evidence through repetition. We adopt the state abstraction practice from the area of automated GUI testing~\citep{crawljax,fraggen} to organize observations into a compact state-transition graph. Each snapshot is normalized by removing script and style tags and mutable component attributes, and replacing timestamps and clock strings with fixed placeholders. Pages with identical normalized representations form one state, and user actions define transitions. Although simple and heuristic, this approach is demonstrated to be effective~\citep{judge}. The constructed state-transition graph provides a straightforward and concise understanding of the application, serving as part of the evidence for scoring.

\subsection{Scoring}
\label{sec:eval-score}

WebCraftBench scores visual aesthetics, usability, and requirement alignment separately on a 0--100 scale. Appendix~\ref{app:scoring-details} provides procedural details.

\textbf{Visual aesthetics.} This dimension assesses up to five representative screenshots collected during exploration. We filter blank and partially loaded frames using foreground and edge density, then apply farthest-point sampling to select visually diverse screenshots. The judges examine layout, information organization, typography, graphic quality, and polish.

\textbf{Usability.} This dimension assesses the experience of interacting with the application using the complete exploration trace and constructed state-transition graph. The judges examine successfully executed interactions and observed failures, including problems with feedback, navigation, state recovery, and fault tolerance.

Both dimensions use a three-round deliberative protocol: an advocate identifies strengths, a critic identifies defects, and a final judge verifies both assessments against the supplied evidence before scoring. The advocate and critic cite concrete observations without assigning scores. The judge adds seven dimension-specific adjustments to a 0--5 base score and clips the result to $[0,5]$. Appendix~\ref{app:scoring-details} specifies these aspects. To reduce sampling variance, the final judge runs independently five times with the first two assessments held fixed. Given scores $q_1,\ldots,q_5\in[0,5]$, we discard the highest and lowest scores and map the remaining mean to 0--100:
\[
s=20\times\frac{\sum_{i=1}^{5}q_i-\max_i q_i-\min_i q_i}{3}.
\]
This protocol aims to improve score discrimination and reduce concentration at the scale endpoints; Appendix~\ref{app:three-round-comparison} compares it with single-round scoring.

\textbf{Requirement alignment.} This dimension judges the application behavior against the acceptance criterion. Since the complete set of evidence can overwhelm an LLM, we build a judging agent for this dimension. This agent starts with the state-transition graph to form an overall understanding of the application. It then carefully examines and searches through recorded evidence using a tool set, which supports regular-expression searches over the DOMs and traces with line-numbered results, paginated reading of DOMs and traces, screenshot inspection, and state-transition graph inspection.
This retrieval process allows the agent to examine relevant evidence without loading the entire record into one context window. 
Once this agent determines that it has collected enough evidence for scoring, it gives a binary judgment on whether each acceptance criterion is implemented.
Every judgment must cite runtime evidence, such as a trace step or a state's DOM snapshot and screenshot.
Each alignment sub-dimension is scored as 100 times the fraction of its criteria judged to be implemented.

\textbf{Overall score.}
Because scores differ in discriminative power across dimensions, we compute z-scores for each dimension to aggregate them while preventing any one dimension from dominating the overall score. Specifically, for model $m$ and dimension $j$, let $\bar{x}_{m,j}$ be the mean raw score over generated applications. We compute
\[
z_{m,j}=\frac{\bar{x}_{m,j}-\mu_j}{\sigma_j},
\]
where $\mu_j$ and $\sigma_j$ are the mean and population standard deviation of these model-level means across the $N$ evaluated LLMs. The mean of the functional, content, and visual alignment $z$-scores gives requirement alignment. Averaging requirement alignment, aesthetics, and usability yields the overall score, with weights of $1/3$ for aesthetics and usability and $1/9$ for each alignment sub-dimension. Scores express relative performance within the evaluated model pool.

\section{Evaluation Results}
\label{sec:results}

Our evaluation aims to answer five research questions.


\subsection{Evaluation Setup}
\label{sec:results-settings}

We evaluate 17 frontier LLMs on all \numtasks{} real-world user requirements. Each model generates one application per requirement through an agentic coding framework, yielding $17\times369=6{,}273$ applications and reflecting practical workflows for web application generation. To make the fullest possible use of model capabilities and reflect practical usage, we integrate the GPT-family LLMs with Codex, and the remaining LLMs with Claude Code. Table~\ref{tab:model-results} lists the reasoning-effort settings; other parameters retain the provider defaults of each model. All applications undergo the same four-stage evaluation pipeline (\S\ref{sec:eval}), without model-specific adjustments. Unless otherwise stated, Claude-Opus-4.8 performs both exploration and scoring.

\subsection{RQ1: Main Results}
\label{sec:results-models}

Table~\ref{tab:model-results} reports dimension-level $z$-scores and the aggregated main scores across 17 LLMs. Positive scores indicate performance above the mean value of the model pool (\S\ref{sec:eval-score}). Appendix~\ref{app:raw} reports detailed raw scores.

\begin{table}[t]
\caption{Results of 17 frontier LLMs on WebCraftBench.}
\label{tab:model-results}
\resizebox{\textwidth}{!}{
\begin{tabular}{rllcccccccc}
\toprule
& & & & \multicolumn{2}{c}{\textbf{General}} & \multicolumn{4}{c}{\textbf{Requirement Alignment}} & \\
\cmidrule(lr){5-6} \cmidrule(lr){7-10}
\textbf{\#} & \textbf{Model} & \textbf{Har.} & \textbf{Eff.} & \textbf{Aesth.} & \textbf{Usab.} & \textbf{Func.} & \textbf{Cont.} & \textbf{Vis.} & \textbf{Align.} & \textbf{Total} \\
\midrule
1  & Claude-Opus-5        & CC & xhigh    & \zposcell{16}{+1.428} & \zposcell{19}{+1.757} & \zposcell{13}{+1.093} & \zposcell{10}{+0.780} & \zposcell{8}{+0.509} & \zposcell{11}{+0.794} & \zposcell{15}{\mathbf{+1.326}} \\
2  & GPT-5.6-Sol          & CX & xhigh    & \zposcell{23}{+2.239} & \zposcell{11}{+0.854} & \zposcell{9}{+0.612} & \zposcell{5}{+0.092} & \znegcell{6}{-0.294} & \zposcell{5}{+0.137} & \zposcell{13}{+1.077} \\
3  & Qwen3.8-Max          & CC & enabled  & \zposcell{11}{+0.792} & \zposcell{13}{+1.088} & \zposcell{12}{+1.008} & \zposcell{13}{+1.049} & \zposcell{13}{+1.035} & \zposcell{13}{+1.031} & \zposcell{12}{+0.970} \\
4  & Kimi-K3              & CC & max      & \zposcell{9}{+0.591} & \zposcell{13}{+1.027} & \zposcell{13}{+1.139} & \zposcell{14}{+1.252} & \zposcell{16}{+1.460} & \zposcell{15}{+1.283} & \zposcell{12}{+0.967} \\
5  & Hy4 preview          & CC & high     & \zposcell{8}{+0.505} & \zposcell{13}{+1.099} & \zposcell{12}{+0.937} & \zposcell{15}{+1.313} & \zposcell{13}{+1.134} & \zposcell{13}{+1.128} & \zposcell{12}{+0.911} \\
6  & Claude-Opus-4.8      & CC & xhigh    & \zposcell{5}{+0.158} & \zposcell{11}{+0.848} & \zposcell{16}{+1.405} & \zposcell{13}{+1.107} & \zposcell{15}{+1.265} & \zposcell{14}{+1.259} & \zposcell{10}{+0.755} \\
7  & Grok-4.5             & CC & high     & \zposcell{8}{+0.527} & \zposcell{12}{+0.902} & \zposcell{12}{+0.923} & \zposcell{9}{+0.648} & \zposcell{11}{+0.890} & \zposcell{11}{+0.820} & \zposcell{10}{+0.750} \\
8  & GPT-5.5              & CX & xhigh    & \zposcell{10}{+0.729} & \zposcell{6}{+0.225} & \zposcell{7}{+0.318} & \zposcell{8}{+0.518} & \zposcell{8}{+0.513} & \zposcell{8}{+0.450} & \zposcell{8}{+0.468} \\
9  & Claude-Opus-4.7      & CC & xhigh    & \znegcell{5}{-0.157} & \znegcell{8}{-0.442} & \znegcell{6}{-0.251} & \znegcell{5}{-0.163} & \znegcell{5}{-0.149} & \znegcell{6}{-0.188} & \znegcell{6}{-0.262} \\
10 & GLM-5.2              & CC & max      & \znegcell{6}{-0.258} & \znegcell{9}{-0.552} & \znegcell{8}{-0.443} & \zposcell{5}{+0.122} & \znegcell{5}{-0.081} & \znegcell{5}{-0.134} & \znegcell{7}{-0.315} \\
11 & DeepSeek-V4-Flash    & CC & max      & \znegcell{7}{-0.312} & \znegcell{7}{-0.322} & \znegcell{8}{-0.479} & \znegcell{7}{-0.374} & \znegcell{7}{-0.369} & \znegcell{7}{-0.407} & \znegcell{7}{-0.347} \\
12 & Hy3                  & CC & high     & \znegcell{10}{-0.665} & \znegcell{8}{-0.519} & \znegcell{10}{-0.673} & \znegcell{10}{-0.725} & \znegcell{7}{-0.379} & \znegcell{9}{-0.592} & \znegcell{9}{-0.592} \\
13 & Qwen3.7-Max          & CC & enabled  & \znegcell{12}{-1.014} & \znegcell{12}{-0.976} & \znegcell{10}{-0.705} & \znegcell{13}{-1.052} & \znegcell{8}{-0.435} & \znegcell{10}{-0.731} & \znegcell{12}{-0.907} \\
14 & GLM-5.1              & CC & enabled  & \znegcell{14}{-1.145} & \znegcell{13}{-1.022} & \znegcell{9}{-0.597} & \znegcell{9}{-0.655} & \znegcell{10}{-0.733} & \znegcell{10}{-0.662} & \znegcell{12}{-0.943} \\
15 & DeepSeek-V4-Pro-Prev & CC & max      & \znegcell{16}{-1.453} & \znegcell{15}{-1.348} & \znegcell{11}{-0.786} & \znegcell{9}{-0.576} & \znegcell{7}{-0.391} & \znegcell{9}{-0.584} & \znegcell{13}{-1.128} \\
16 & Kimi-K2.7-Code       & CC & enabled  & \znegcell{18}{-1.642} & \znegcell{14}{-1.176} & \znegcell{13}{-1.129} & \znegcell{10}{-0.685} & \znegcell{17}{-1.529} & \znegcell{13}{-1.114} & \znegcell{15}{-1.311} \\
17 & MiniMax-M3           & CC & adaptive & \znegcell{7}{-0.323} & \znegcell{16}{-1.443} & \znegcell{24}{-2.371} & \znegcell{26}{-2.652} & \znegcell{24}{-2.447} & \znegcell{25}{-2.490} & \znegcell{16}{-1.419} \\
\bottomrule
\end{tabular}}
{\footnotesize
\textbf{Note:} Har.: coding harness (CX: Codex; CC: Claude Code). Eff.: reasoning effort.
}
\end{table}

\emph{No model leads on every dimension.} Claude-Opus-5 ranks first overall and leads in usability ($+1.757$); GPT-5.6-Sol leads in aesthetics by a large margin ($+2.239$), while Kimi-K3 leads in requirement alignment ($+1.283$). These complementary strengths would be obscured by an overall ranking alone.

\emph{Visual aesthetics and usability capture distinct aspects of application quality.} Their correlation is strong at the model level ($r=0.85$) but only $0.36$ at the application level. GPT-5.6-Sol ranks first in aesthetics and ranks sixth in usability; MiniMax-M3 has near-average aesthetics ($-0.323$) but the lowest usability ($-1.443$). Visual polish therefore does not establish reliable interaction, supporting runtime evaluation alongside visual assessment.

\emph{Model generations differ under the tested configurations.} Within each family represented by multiple generations, the newer generation has a higher overall score, indicating steady progress in the capabilities of each model family.

\subsection{RQ2: Human Preference Alignment}
\label{sec:results-agreement}

We sample 200 sessions from our internal replica of Code Arena~\citep{WebDevArena}, independently of the 369 benchmark tasks. Among these sessions, 197 retain valid application pairs. Each session records a human vote (preference) between two applications for the same requirement. In order to eliminate irresponsible votes, six additional annotators review and fix every vote, and after that, two others further review a random 20\% subset independently. The validated votes serve as ground truth labels.

To obtain WebCraftBench's automated votes, for application $a$ and component $j$, we compute $z_{a,j}=(x_{a,j}-\mu_j^{H})/\sigma_j^{H}$ from its raw score $x_{a,j}$, using the component mean $\mu_j^{H}$ and standard deviation $\sigma_j^{H}$ over all applications in this human-agreement experiment. We aggregate these $z$-scores using the same process as in \S\ref{sec:eval-score} and compare the resulting pairwise ordering with human preferences. We define $\Delta z$ as the absolute difference between paired overall scores and report within-band agreement and cumulative agreement above each threshold.

\begin{table}[t]
\caption{Agreement between WebCraftBench and human pairwise preferences.}
\label{tab:agreement}
\centering
\resizebox{0.65\textwidth}{!}{
\begin{tabular}{lrrrrrr}
\toprule
& \multicolumn{3}{c}{\textbf{Within band}} & \multicolumn{3}{c}{\textbf{Cumulative} ($\Delta z \geq$ lower bound)} \\
\cmidrule(lr){2-4} \cmidrule(lr){5-7}
\textbf{$\Delta z$ band} & \textbf{$n$} & \textbf{Agree} & \textbf{Rate} & \textbf{$n$} & \textbf{Agree} & \textbf{Rate} \\
\midrule
$[0,\ 0.25)$     & 65 & 48 & 73.8\%  & 197 & 168 & \textbf{85.3\%} \\
$[0.25,\ 0.5)$   & 52 & 45 & 86.5\%  & 132 & 120 & 90.9\% \\
$[0.5,\ 0.75)$   & 30 & 26 & 86.7\%  & 80  & 75  & 93.8\% \\
$[0.75,\ 1.0)$   & 13 & 13 & 100.0\% & 50  & 49  & 98.0\% \\
$[1.0,\ 1.5)$    & 17 & 16 & 94.1\%  & 37  & 36  & 97.3\% \\
$[1.5,\ 2.0)$    & 7  & 7  & 100.0\% & 20  & 20  & \textbf{100.0\%} \\
$[2.0,\ \infty)$ & 13 & 13 & 100.0\% & 13  & 13  & 100.0\% \\
\bottomrule
\end{tabular}}
\end{table}

Table~\ref{tab:agreement} shows that WebCraftBench aligns with human preferences on 168 of 197 pairs, yielding a satisfactory overall agreement rate of 85.3\%. 

Disagreements are concentrated among pairs with small score differences. Of the 29 disagreements, 17 occur in the $[0,0.25)$ band and 7 in $[0.25,0.5)$. Cumulative agreement exceeds 90\% for pairs with $\Delta z\geq0.25$. Larger score differences are generally associated with higher agreement. Among the 50 pairs with $\Delta z\geq0.75$, the only disagreement involves a tradeoff between visual quality and functional reliability: one application has a polished interface but crashes on its primary functional path, while the other has a less polished interface and working core functionality. WebCraftBench prefers the latter, whereas the human rater prefers the former. Appendix~\ref{app:agreement-case} presents the applications and examines this disagreement.

\subsection{RQ3: Judge Robustness}
\label{sec:results-judge}

In order to test whether WebCraftBench depends on the capabilities of a particular LLM and to prevent model-family bias, we replace the main judge throughout scoring with Gemini-3.7-Flash, a lower-cost model from a different provider, while holding tasks, generated applications, exploration traces, procedures, and prompts fixed. Overall scores have a Pearson correlation~\citep{pearson} of $r=0.983$ and a Spearman rank correlation~\citep{spearman} of $\rho=0.980$ across judges. Of the $\binom{17}{2}=136$ model pairs, 130 retain their relative order (95.6\% agreement). Three of the six reversals involve main-judge score differences below 0.04. Although raw-score levels and spreads change substantially, aggregate rankings remain stable. For example, the model-pool mean usability score increases from 60.1 to 75.7, while the range of model-level aesthetics scores expands from 26.8 to 44.2 points. These changes motivate standardizing dimensional raw scores before aggregation. Appendix~\ref{app:judge-details} provides the full ranking comparison and raw-score analysis.

\subsection{RQ4: Instrumentation Reliability}
\label{sec:results-instrumentation}

We evaluate instrumentation applicability across all 6{,}273 applications in the main experiment. Instrumentation fails for only 7 applications, yielding a success rate of 99.89\%. The failures have identifiable engineering causes: 4 applications use JSX input forms unsupported by Istanbul, 2 use npm-workspaces monorepos for which the tool cannot locate the Vite project root, and 1 exceeds the sandbox deployment-size limit after instrumentation. We separately check deployment regressions on the 360 deployable applications generated by Claude-Opus-5. All remain deployable after instrumentation. This check establishes deployment reliability on the tested subset. Under the fallback procedure in \S\ref{sec:eval-instrument}, the 7 applications for which instrumentation fails still undergo exploration and scoring in black-box mode, without coverage guidance.

\subsection{RQ5: Coverage Improvement}
\label{sec:results-exploration}

\begin{figure}[t]
\begin{center}
\includegraphics[width=\textwidth]{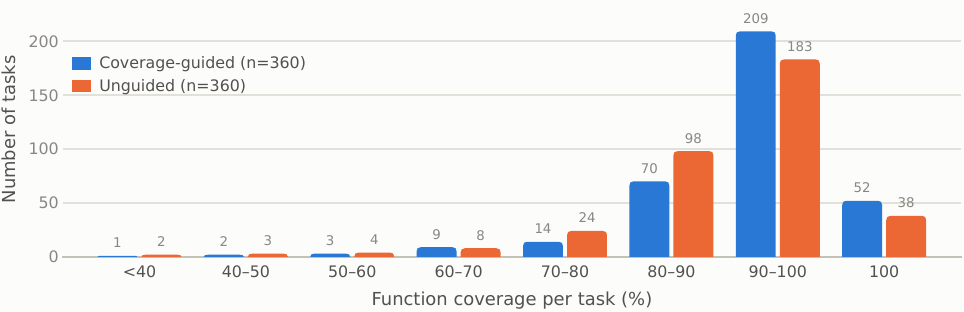}
\end{center}
\caption{Effect of coverage guidance on function coverage during exploration.}
\label{fig:coverage}
\end{figure}

Code coverage measures the proportion of the implementation exercised during exploration and helps identify functionality that may remain unobserved. We assess the contribution of coverage guidance by comparing exploration with and without this feedback. To control experimental scale, this ablation uses the 360 deployable applications generated by Claude-Opus-5. The remaining 9 applications fail for generation-side reasons, including missing files, deployment errors, and incorrect port declarations.

Figure~\ref{fig:coverage} compares the coverage distributions under the two settings. Coverage guidance increases median function coverage from 91.9\% to 94.3\%. The number of tasks below 90\% coverage decreases from 139 (39\%) to 99 (28\%), with a relative reduction of 29\%, while the number reaching 100\% coverage increases from 38 to 52. The gains are concentrated in the medium-to-high coverage range: the $[70\%,90\%)$ interval contains 38 fewer tasks, and the number at or above 90\% increases by 40, from 221 to 261. The low-coverage tail changes little, with 17 tasks below 70\% coverage without guidance and 15 with guidance.

Manual inspection shows that applications below 70\% coverage typically contain failures that block further exploration. Examples include an entry page that immediately reports an error, an event handler that throws an exception, and a broken critical path that makes downstream functionality unreachable. These observations suggest that application defects contribute substantially to the remaining low coverage. Such failures also prevent real users from reaching the affected functionality, making them relevant to the evaluation of user experience.

In addition, the LLM-generated applications remain simple. Prior GUI-testing studies~\citep{empirical, webrled} report greater exploration difficulty on complex applications~\citep{temac,webrled}; coverage guidance is expected to bring more benefits as model capabilities grow and models generate more complex applications.

\section{Limitations}
\label{sec:limitations}

\textbf{LLM judges.} The evaluation process largely relies on LLMs. Therefore, judge preferences may affect the results. We validated the reliability of the scores by comparing them with human preferences (\S\ref{sec:results-agreement}) and verified scoring stability by comparing results after replacing the judge (\S\ref{sec:results-judge}).

\textbf{Model-pool dependence.} Standardization makes $z$-scores dependent on the participating LLMs and unsuitable for direct comparison across pools. Future deployments could use fixed reference means and standard deviations. Raw scores further aid interpretation.

\textbf{Task scope.} The dataset covers single-turn, text-only, front-end-only tasks from one platform, and excludes iterative repair, multimodal input, and back-end integration. Since the evaluation process simulates how real users use webpages, it can be extended to other frameworks and webpages with back-end functionality; considering the engineering cost, we leave this as future work.

\textbf{Open source.} To prevent data leakage and comply with our business confidentiality policy, we do not publicly release the dataset or source code. 

\section{Conclusion}

In this paper, we introduced WebCraftBench, which evaluates web application generation from a software testing perspective. Combining coverage-guided exploration, state abstraction, and separated scoring grounded in runtime evidence, WebCraftBench can objectively and comprehensively score generated web applications, thereby reliably evaluating the capability of LLMs to generate web applications. WebCraftBench achieves 85.3\% agreement with human preferences, with stable rankings across judge models. Coverage guidance increases the coverage during exploration, while the results across 17 frontier LLMs reveal complementary strengths in aesthetics, usability, and requirement alignment.

\bibliography{ref}
\bibliographystyle{templates/references}

\clearpage
\setcounter{page}{1}
\appendix

\begin{center}
{\large\bfseries Appendices}
\end{center}

\textbf{\large Table of Contents:}

\startcontents[appendix]
\printcontents[appendix]{l}{1}{}

\newpage

\section{Evaluation Details}
\label{app:eval-details}

\subsection{Instrumentation and Deployment}
\label{app:instrumentation-details}

The tool described in \S\ref{sec:eval-instrument} instruments static and plain Node.js projects directly at the source level. For projects requiring compilation, it injects an instrumentation plugin and processes the generated files after the build. Framework detection selects the corresponding procedure automatically, and the instrumented application is deployed in an isolated sandbox over HTTP.

Instrumentation depends on application toolchains and may affect deployment or runtime behavior. Section~\ref{sec:results-instrumentation} reports its applicability and deployment checks. Since Istanbul is a time-tested and widely used instrumentation tool, we believe that the instrumentation does not harm the application behavior. The black-box fallback retains applications with instrumentation failures for exploration and scoring without coverage feedback.

\subsection{Scoring Procedures}
\label{app:scoring-details}

This section details the evidence requirements, judgment procedures, and scoring aspects used in \S\ref{sec:eval-score}.

\textbf{Deliberation and score aggregation.} Visual aesthetics and usability each use three rounds. The advocate identifies strengths, and the critic identifies defects; both must cite concrete observations without assigning scores. The final judge checks their claims against the supplied evidence, discards unsupported claims, and adds relevant observations that both rounds missed. It selects a 0--5 base score from the overall visual impression or interaction experience, then records a signed adjustment and a supporting rationale for each of seven aspects, explicitly indicating when no adjustment is needed. The sum of the base score and adjustments is clipped to $[0,5]$. With the advocate and critic assessments held fixed, the final judge runs independently five times. The highest and lowest scores are discarded, and the remaining mean is multiplied by 20 to obtain the 0--100 score.

\textbf{Visual aesthetics.} The seven aspects are first impression, content completeness, layout structure, visual detail, information hierarchy and typography, stylistic consistency, and emotion and trust. Observations must refer to specific locations in the representative screenshots. The assessment distinguishes deliberate simplicity from unfinished or empty content and examines the quality of graphics and layout rather than merely their presence. Factual correctness alone does not establish visual quality, and the judge must not infer failures in unobserved dynamic behavior from static screenshots.

\textbf{Usability.} The seven interaction-oriented aspects are effectiveness of core operations, immediacy of feedback, quality of controls and input, interaction flow and discoverability, state reachability and recovery, avoidance of dead ends and errors together with response consistency, and stability and fault tolerance. Each assessment must cite an exploration-trace step or a state-transition graph node or edge. The judge considers observed responses, navigation, recovery, and failures, without inventing requirements for unobserved features. A path that was not tested is neither a strength nor a weakness; observed failures that block core operations warrant substantial penalties.

\textbf{Requirement alignment.} Functional, content, and visual criteria share one evidence-retrieval process but receive separate satisfaction percentages. Starting from the state-transition graph and state overview, the agent retrieves relevant trace steps, DOM passages, and screenshots as needed. Functional criteria are checked against available features, executed interactions, and resulting state changes; content criteria against the specified text, data, and resources; and visual criteria against the required appearance and components. The agent reports a Boolean decision, supporting evidence, a brief analysis, and a confidence estimate for each criterion. A positive decision requires evidence that the property is present in the running application; required functionality must be reachable and observed to work. Unexecuted source code alone is insufficient, and no partial-credit category is used. Each sub-dimension receives 100 times the fraction of its criteria judged satisfied.

\newpage

\section{Detailed Judge-Robustness Results}
\label{app:judge-details}
The experiment in \S\ref{sec:results-judge} replaces Claude-Opus-4.8 with Gemini-3.7-Flash while holding the generated applications, exploration traces, scoring procedures, and prompts fixed. We report the complete ranking comparison and changes in raw-score distributions below.

Table~\ref{tab:judge-ablation} shows that model rankings remain stable across the two judges. Overall model scores have a Pearson correlation of $r=0.983$ and a Spearman rank correlation of $\rho=0.980$. Of the 136 model pairs, 130 retain their relative order, corresponding to 95.6\% pairwise agreement. The top four ranks and ranks 8, 9, and 12 remain unchanged.

Of the six pairs whose order changes, three have overall-score differences below 0.04 under the main judge. This pattern is consistent with the concentration of human disagreements among near-tied applications in \S\ref{sec:results-agreement}. Of the remaining three reversals, one is between Hy4 preview and Grok-4.5, while two involve MiniMax-M3, which rises from rank 17 to 15 under Gemini-3.7-Flash. Hy4 preview moves from rank 5 to 6, and Grok-4.5 moves from rank 7 to 5. All models move by at most two rank positions.

Raw-score changes vary by dimension; Appendix~\ref{app:raw} reports the underlying means and ranges. Scores are standardized separately within each judge's 17-model pool and aggregated with the same weights. The correlations and pairwise agreement above quantify the resulting ranking robustness.

\begin{table}[H]
\caption{Score and ranking differences across 17 models. Main: Claude-Opus-4.8; Gem.: Gemini-3.7-Flash. Scores are standardized within the 17-model pool. Rk. is the Gemini rank; $\Delta$ is that rank minus the main-judge rank.}
\label{tab:judge-ablation}
\centering
\footnotesize
\setlength{\tabcolsep}{3.5pt}
\resizebox{\textwidth}{!}{%
\begin{tabular}{rlrrrc@{\hskip 1.5em}rlrrrc}
\toprule
& & \multicolumn{2}{c}{\textbf{Total}} & & & & & \multicolumn{2}{c}{\textbf{Total}} & & \\
\cmidrule(lr){3-4} \cmidrule(lr){9-10}
\textbf{\#} & \textbf{Model} & \textbf{Main} & \textbf{Gem.} & \textbf{Rk.} & \textbf{$\Delta$} &
\textbf{\#} & \textbf{Model} & \textbf{Main} & \textbf{Gem.} & \textbf{Rk.} & \textbf{$\Delta$} \\
\midrule
1 & Claude-Opus-5 & $+1.326$ & $+1.602$ & 1 & 0 & 10 & GLM-5.2 & $-0.315$ & $-0.254$ & 11 & $+1$ \\
2 & GPT-5.6-Sol & $+1.077$ & $+1.331$ & 2 & 0 & 11 & DeepSeek-V4-Flash & $-0.347$ & $-0.107$ & 10 & $-1$ \\
3 & Qwen3.8-Max & $+0.970$ & $+0.816$ & 3 & 0 & 12 & Hy3 & $-0.592$ & $-0.796$ & 12 & 0 \\
4 & Kimi-K3 & $+0.967$ & $+0.729$ & 4 & 0 & 13 & Qwen3.7-Max & $-0.907$ & $-0.987$ & 14 & $+1$ \\
5 & Hy4 preview & $+0.911$ & $+0.712$ & 6 & $+1$ & 14 & GLM-5.1 & $-0.943$ & $-0.944$ & 13 & $-1$ \\
6 & Claude-Opus-4.8 & $+0.755$ & $+0.671$ & 7 & $+1$ & 15 & DeepSeek-V4-Pro-Prev & $-1.128$ & $-1.206$ & 16 & $+1$ \\
7 & Grok-4.5 & $+0.750$ & $+0.726$ & 5 & $-2$ & 16 & Kimi-K2.7-Code & $-1.311$ & $-1.413$ & 17 & $+1$ \\
8 & GPT-5.5 & $+0.468$ & $+0.357$ & 8 & 0 & 17 & MiniMax-M3 & $-1.419$ & $-1.172$ & 15 & $-2$ \\
9 & Claude-Opus-4.7 & $-0.262$ & $-0.067$ & 9 & 0 & \multicolumn{6}{c}{} \\
\bottomrule
\end{tabular}}

\end{table}

\newpage

\section{The High-Margin Human--WebCraftBench Disagreement Case}
\label{app:agreement-case}

\begin{figure}[H]
\centering
\begin{minipage}[t]{0.49\textwidth}
\centering
\includegraphics[width=\linewidth]{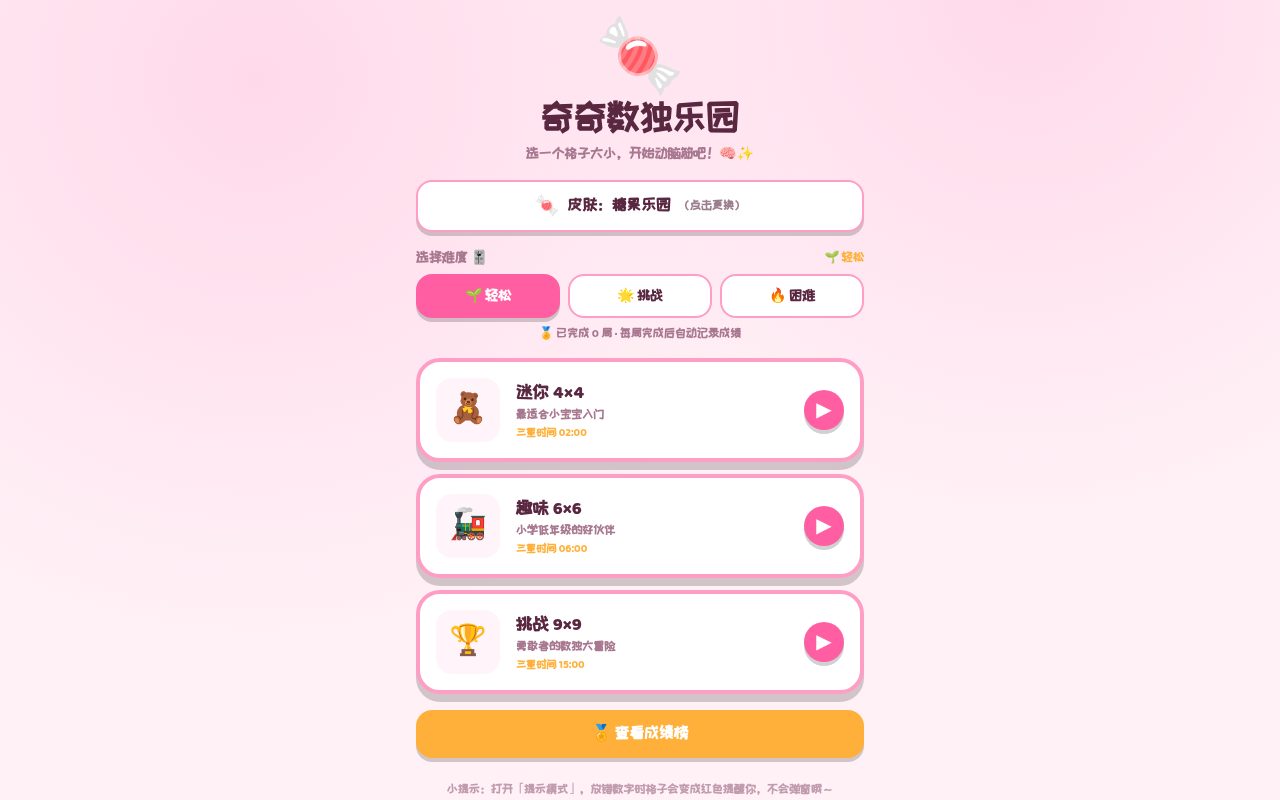}\\[-2pt]
\small (a) Human-preferred artifact: more visually polished, but the primary gameplay path crashes.
\end{minipage}
\hfill
\begin{minipage}[t]{0.49\textwidth}
\centering
\includegraphics[width=\linewidth]{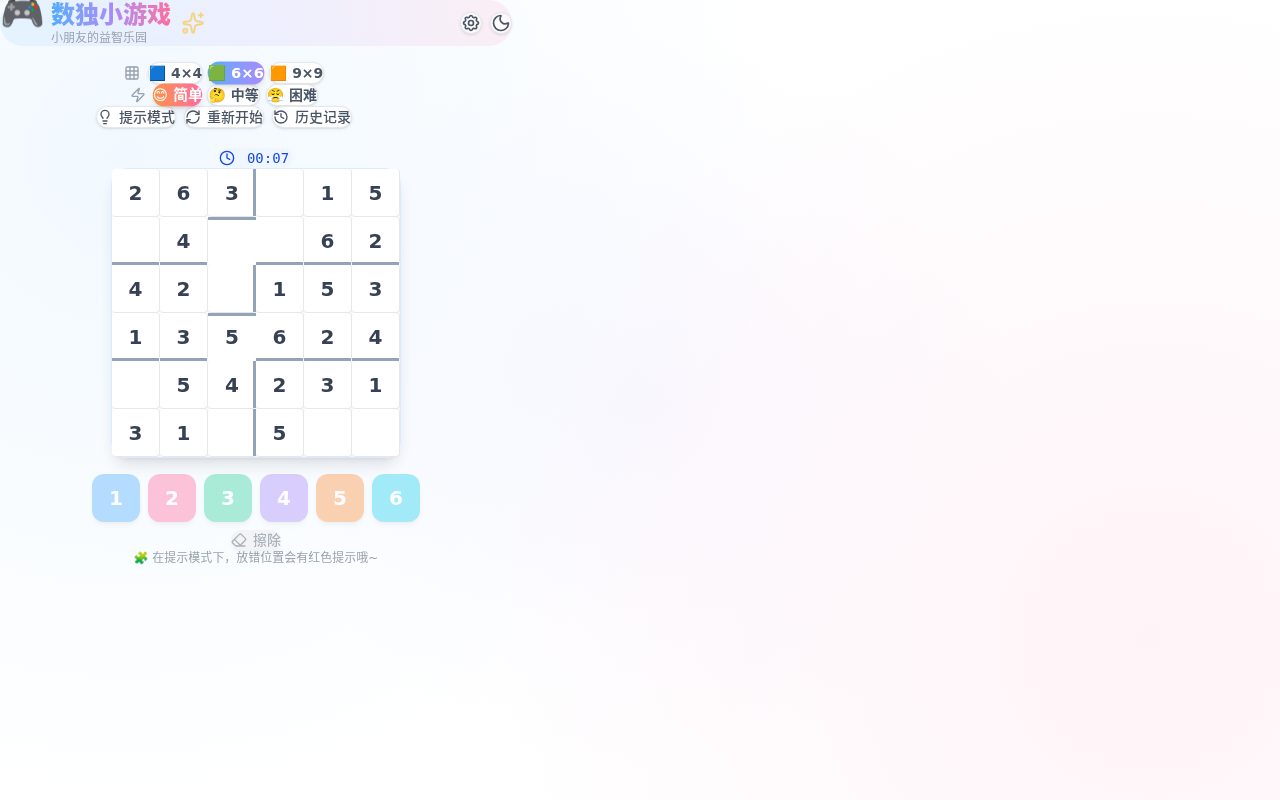}\\[-2pt]
\small (b) WebCraftBench-preferred artifact: less visually polished, but the core Sudoku interaction works.
\end{minipage}
\caption{The only disagreement among the 50 human-preference pairs with $\Delta z\geq0.75$. The human rater prefers the visually polished artifact in (a); WebCraftBench prefers the artifact in (b), whose core functionality works during runtime testing.}
\label{fig:agreement-case}
\end{figure}

Figure~\ref{fig:agreement-case} shows the only disagreement among the 50 human-preference pairs with $\Delta z\geq0.75$. The human rater prefers the artifact in Figure~\ref{fig:agreement-case}(a), which has a more polished and visually coherent page. Runtime testing reveals that its primary gameplay path crashes after entry. The artifact in Figure~\ref{fig:agreement-case}(b) has a sparser interface, but its core Sudoku interaction works. WebCraftBench prefers this artifact because its advantages in usability and requirement alignment outweigh the visual advantage of the first artifact. The case is consistent with different tradeoffs between aesthetics and functional reliability, although the vote alone does not identify the rater's weighting.

\newpage

\section{Three-Round versus Single-Round Scoring}
\label{app:three-round-comparison}

\begin{figure}[H]
\centering
\includegraphics[width=\textwidth]{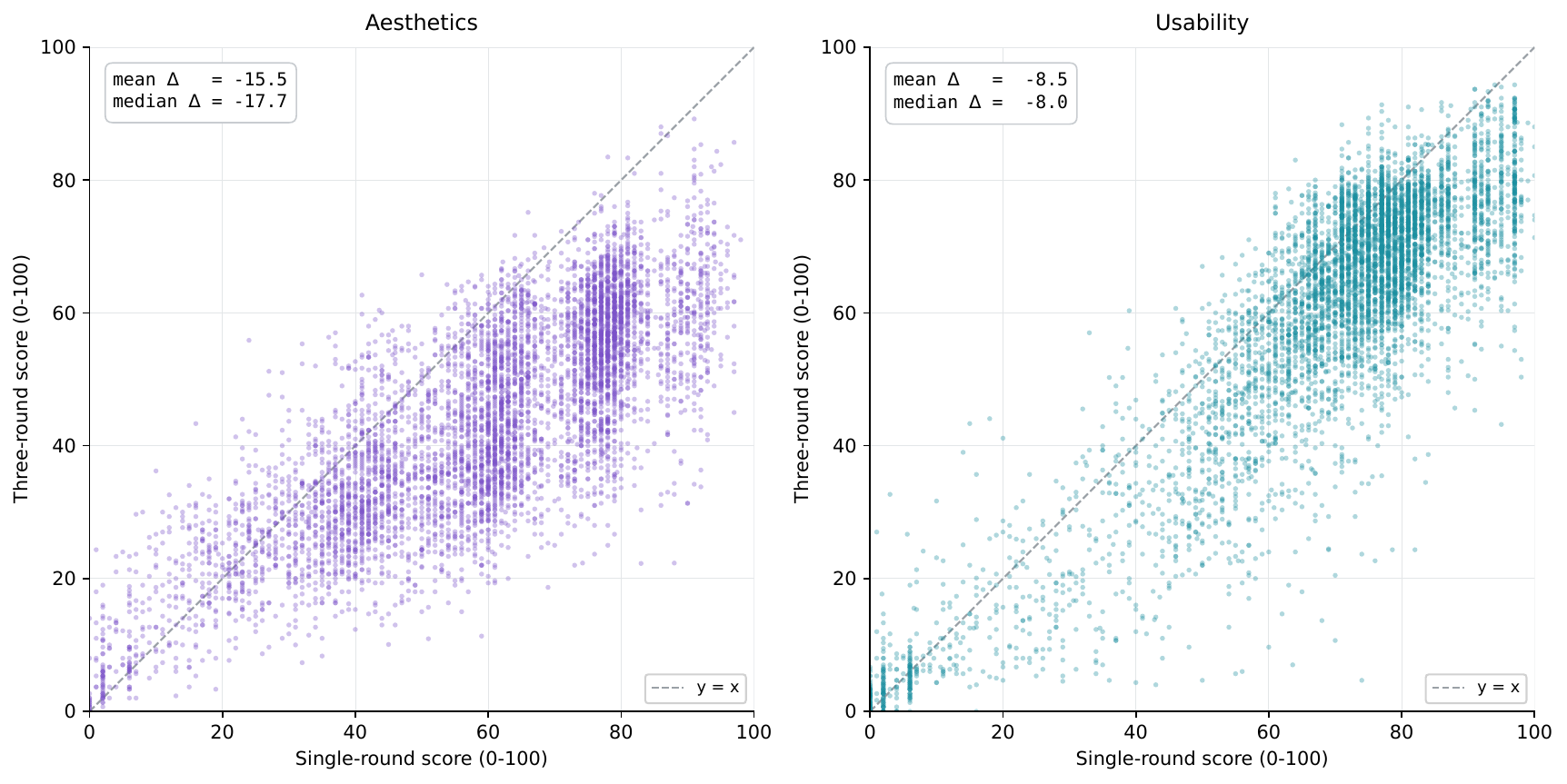}
\caption{Application-level comparison of direct single-round scoring (horizontal axis) and the three-round deliberative protocol used by WebCraftBench (vertical axis) for visual aesthetics and usability. Each point is one paired evaluation; the dashed line marks equal scores. Here, $\Delta$ denotes the three-round score minus the single-round score.}
\label{fig:three-round-comparison}
\end{figure}

We compare application-level scores from the three-round deliberative protocol with those from direct single-round scoring to assess the effect on score distributions. The single-round baseline directly uses the final round prompt of the three-round protocol. Figure~\ref{fig:three-round-comparison} presents paired results for visual aesthetics and usability, with the dashed diagonal indicating equal scores. The deliberative protocol lowers mean scores by 15.5 points for aesthetics and 8.5 points for usability, suggesting that advocacy, criticism, and evidence verification lead to more conservative judgments. Artifacts assigned identical or similarly saturated scores in a single round also receive a wider range of scores after deliberation. These results indicate reduced concentration at the scale endpoints and greater differentiation among artifacts that direct scoring evaluates similarly. 

\newpage

\section{Raw Scores}
\label{app:raw}

The main-text rankings aggregate standardized component scores to account for differences in component dispersion (\S\ref{sec:eval-score}). Scores are relative to the evaluated model pool and are not directly comparable across studies. This appendix reports raw means under Claude-Opus-4.8 and Gemini-3.7-Flash for the same 17-model pool used in \S\ref{app:judge-details}.

Table~\ref{tab:raw-scores} reports raw scores on a 0--100 scale. For visual aesthetics and usability, five final-judge scores on a 0--5 scale are aggregated using a trimmed mean and multiplied by 20. Raw means are rounded to one decimal place for presentation and may obscure distinctions retained in the standardized scores. Rows follow Table~\ref{tab:model-results}. Column means and ranges are computed over all 17 models, consistently with the ranking comparison.

Additional correlations highlight complementary aspects of application quality: visual aesthetics and requirement alignment are weakly correlated at the artifact level ($r=0.32$), while usability and requirement alignment are strongly associated at the model level ($r=0.88$).

Component scales differ substantially: under the main judge, aesthetics spans 26.8 points across models, whereas the alignment sub-dimensions span 7.1--10.4 points. Averaging raw components would therefore give aesthetics greater influence on between-model differences. Standardization equalizes component dispersion before applying the stated weights. The cross-judge differences in this table concern score level and spread; ranking robustness is evaluated separately in Table~\ref{tab:judge-ablation}.

\begin{table}[H]
\caption{Raw model-level mean scores under both judges (0--100). Alignment scores are percentages of criteria satisfied. Rows follow Table~\ref{tab:model-results}; column means and ranges cover all 17 models.}
\label{tab:raw-scores}
\centering
\footnotesize
\setlength{\tabcolsep}{4pt}
\resizebox{\textwidth}{!}{%
\begin{tabular}{rl rr rrr @{\hskip 1.5em} rr rrr}
\toprule
& & \multicolumn{5}{c}{\textbf{Main judge} (Claude-Opus-4.8)} & \multicolumn{5}{c}{\textbf{Gemini-3.7-Flash}} \\
\cmidrule(lr){3-7} \cmidrule(lr){8-12}
& & & & \multicolumn{3}{c}{Alignment} & & & \multicolumn{3}{c}{Alignment} \\
\cmidrule(lr){5-7} \cmidrule(lr){10-12}
\textbf{\#} & \textbf{Model} & \textbf{Aesth.} & \textbf{Usab.} & \textbf{Func.} & \textbf{Cont.} & \textbf{Vis.} & \textbf{Aesth.} & \textbf{Usab.} & \textbf{Func.} & \textbf{Cont.} & \textbf{Vis.} \\
\midrule
1 & Claude-Opus-5 & 53.5 & 69.6 & 96.7 & 95.9 & 95.6 & 59.5 & 89.1 & 99.2 & 98.9 & 99.5 \\
2 & GPT-5.6-Sol & 59.1 & 64.7 & 95.4 & 94.7 & 93.7 & 70.3 & 82.5 & 97.6 & 97.9 & 98.0 \\
3 & Qwen3.8-Max & 49.1 & 66.0 & 96.5 & 96.4 & 96.9 & 50.7 & 84.2 & 98.1 & 97.4 & 98.0 \\
4 & Kimi-K3 & 47.7 & 65.6 & 96.9 & 96.8 & 97.9 & 47.2 & 82.7 & 97.9 & 98.1 & 98.6 \\
5 & Hy4 preview & 47.1 & 66.0 & 96.3 & 96.9 & 97.1 & 46.1 & 81.9 & 98.0 & 98.4 & 98.9 \\
6 & Claude-Opus-4.8 & 44.7 & 64.7 & 97.6 & 96.5 & 97.4 & 43.4 & 82.1 & 98.4 & 98.0 & 99.4 \\
7 & Grok-4.5 & 47.3 & 64.9 & 96.3 & 95.7 & 96.5 & 47.4 & 83.1 & 98.0 & 97.4 & 98.9 \\
8 & GPT-5.5 & 48.7 & 61.3 & 94.6 & 95.5 & 95.6 & 49.3 & 75.2 & 97.2 & 97.6 & 98.5 \\
\midrule
9 & Claude-Opus-4.7 & 42.6 & 57.7 & 93.0 & 94.2 & 94.0 & 41.0 & 72.8 & 97.1 & 97.2 & 98.1 \\
10 & GLM-5.2 & 41.9 & 57.1 & 92.5 & 94.8 & 94.2 & 39.3 & 71.7 & 96.2 & 97.5 & 97.4 \\
11 & DeepSeek-V4-Flash & 41.5 & 58.4 & 92.4 & 93.9 & 93.5 & 39.4 & 75.5 & 96.6 & 96.8 & 97.4 \\
12 & Hy3 & 39.0 & 57.3 & 91.9 & 93.2 & 93.5 & 34.9 & 71.5 & 94.4 & 95.4 & 95.8 \\
13 & Qwen3.7-Max & 36.6 & 54.8 & 91.8 & 92.7 & 93.3 & 32.0 & 67.3 & 95.1 & 95.1 & 96.7 \\
14 & GLM-5.1 & 35.7 & 54.6 & 92.1 & 93.4 & 92.6 & 29.1 & 68.5 & 95.8 & 95.0 & 97.2 \\
15 & DeepSeek-V4-Pro-Prev & 33.6 & 52.8 & 91.6 & 93.5 & 93.4 & 27.0 & 66.5 & 94.8 & 95.2 & 96.4 \\
16 & Kimi-K2.7-Code & 32.3 & 53.7 & 90.6 & 93.3 & 90.7 & 26.1 & 66.1 & 94.9 & 94.8 & 94.5 \\
17 & MiniMax-M3 & 41.4 & 52.3 & 87.2 & 89.8 & 88.5 & 40.4 & 66.2 & 92.8 & 94.6 & 94.0 \\
\midrule
\multicolumn{2}{l}{\emph{pool mean}} & 43.6 & 60.1 & 93.7 & 94.5 & 94.4 & 42.5 & 75.7 & 96.6 & 96.8 & 97.5 \\
\multicolumn{2}{l}{\emph{range}} & 26.8 & 17.3 & 10.4 & 7.1 & 9.4 & 44.2 & 23.0 & 6.4 & 4.3 & 5.5 \\
\bottomrule
\end{tabular}}
\end{table}

%% file: sections/rank_teaser.tex
\begingroup
\setlength{\parskip}{0pt}
\noindent
\begin{minipage}{\linewidth}
  \centering
  \includegraphics[width=\linewidth]{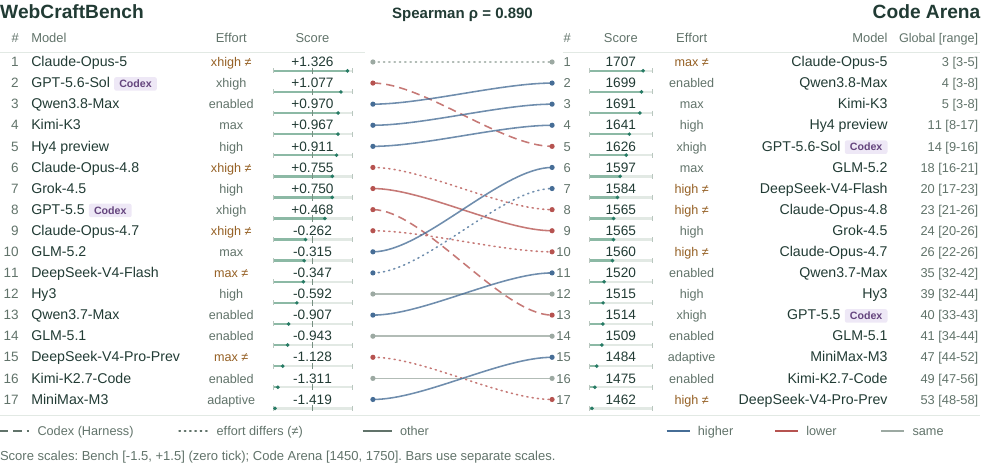}
  \captionsetup{font=footnotesize,skip=4pt,hypcap=false}
  \captionof{figure}{\textbf{WebCraftBench and Code Arena rankings for 17 shared models.}
  \# gives shared-set rank; Global [range] gives Code Arena's overall rank and reported range.
  Blue/red/gray links indicate higher/lower/unchanged shared-set rank.
  Long dashes mark the two GPT models using Codex on both sides; other models use Claude Code in WebCraftBench and a custom harness in Code Arena. Dots mark differing effort labels.
  Scores use separate scales; harness/effort can differ.}
  \label{fig:frontend-alignment}
\end{minipage}
\par
\endgroup
\clearpage